\documentclass[12pt]{article}

\font\grande=cmr9.5 scaled \magstep4
\font\medio=cmr9.5 scaled \magstep2
\outer\def\beginsection#1\par{\medbreak\bigskip
      \message{#1}\leftline{\bf#1}\nobreak\medskip
\vskip-\parskip
      \noindent}
\usepackage{graphicx} 
\begin{document}
\bibliographystyle {unsrt}

\titlepage

\begin{flushright}
CERN-PH-TH/2009-171
\end{flushright}

\vspace{15mm}
\begin{center}
{\grande The V-mode polarization}\\
\vspace{5mm}
{\grande of the Cosmic Microwave Background}\\
\vspace{1.5cm}
 Massimo Giovannini 
 \footnote{Electronic address: massimo.giovannini@cern.ch} \\
\vspace{1cm}
{{\sl Department of Physics, 
Theory Division, CERN, 1211 Geneva 23, Switzerland }}\\
\vspace{0.5cm}
{{\sl INFN, Section of Milan-Bicocca, 20126 Milan, Italy}}
\vspace*{2cm}

\end{center}

\vskip 1cm
\centerline{\medio  Abstract}
The V-mode polarization of the Cosmic Microwave Background is discussed in a weakly magnetized plasma. The VV and VT angular power spectra are computed for adiabatic initial conditions of the Einstein-Boltzmann hierarchy. Depending upon the frequency channel and upon the magnetic field intensity, the VT power spectra of the circular polarization can even be seven orders of magnitude larger than a putative B-mode polarization stemming from the lensing of the primary anisotropies. Specific programs aimed at the direct detection of the V-mode polarization of the Cosmic Microwave Background could provide a new observational tool for the scrutiny of pre-decoupling physics. 
\noindent

\vspace{5mm}

\vfill
\newpage
The spectral energy density (per logarithmic interval of frequency) of the 
Cosmic Microwave Background (CMB in what follows) is maximal, today, for photon energies $E_{\gamma} \sim 9.2\times 10^{-4}$ eV whose associated wavelength $\lambda_{\gamma} \sim 2\pi/E_{\gamma}$ is of the order of $0.13$ mm. According to the WMAP 5-yr data the redshift of hydrogen recombination can be estimated approximately as
$z_{\mathrm{rec}} = 1090.51 \pm 0.95$ \cite{WMAP5} corresponding to the 
conformal time\footnote{The conformal time coordinate $\tau$ will be used throughout; in terms of $\tau$ 
the background metric $\overline{g}_{\mu\nu}$ will be chosen as conformally flat i.e. $\overline{g}_{\mu\nu}= a^2(\tau) \eta_{\mu\nu}$ where 
$\eta_{\mu\nu} = \mathrm{diag}(1,\, -1,\, -1,\,-1)$ is the Minkowski metric.  The current observational 
evidence \cite{WMAP5} suggests, indeed, that the 
spatial curvature can be neglected at the recombination epoch, at least 
in the framework of the concordance model, i.e. the $\Lambda$CDM 
paradigm where $\Lambda$ stands for the dark energy component and CDM stands for the cold dark matter contribution.}
 $\tau_{\mathrm{rec}}$; at $\tau_{\mathrm{rec}}$ the maximum of the CMB is 
 $z_{\mathrm{rec}} E_{\gamma} \simeq \mathrm{eV}$ corresponding to  a physical wavelength 
$\lambda_{\gamma}(\tau_{\mathrm{rec}})=  \lambda_{\gamma}/z_{\mathrm{rec}} \simeq 0.12\,\mu\mathrm{m}$. 
Prior to recombination the electrons and the ions have  kinetic 
temperatures which are comparable with the temperature of the photons, 
i.e. $(1 + z_{\mathrm{rec}}) T_{\gamma 0}$ (where 
$T_{\gamma0} = 2.725\,$K). The small difference between 
electron and proton temperatures is controlled by the ratio between 
the Hubble rate $H$ and the Coulomb rate $\Gamma_{\mathrm{Coul}}$ 
which is ${\mathcal O}(10^{-11})$ at $\tau_{\mathrm{rec}}$. The global 
charge neutrality of the plasma combined with the baryon asymmetry implies 
that the electron and proton concentrations are equal and both 
of the order of $10^{-10}\, n_{\gamma}$ where $n_{\gamma}$ is the comoving
photon concentration. Prior to recombination the plasma is cold: the electron and proton masses 
are both much larger than the kinetic temperature of the corresponding species.  Consider the physical situation when, prior 
to recombination, the plasma is supplemented by a magnetic field whose typical inhomogeneity scale is at least comparable (and possibly even larger) than the Hubble radius $H^{-1}$ at the corresponding epoch. 
Since the wavelengths of the scattered photons are minute in comparison with the Hubble radius  
(i.e. $\lambda_{\gamma}(\tau_{\mathrm{rec}}) \ll H^{-1}$)
 the magnetic field gradients can be ignored, in the first approximation, when computing  the  photon-electron (and photon-ion) scattering.  The gradient expansion on the magnetic field strength was termed long ago 
 by Alfv\'en guiding centre approximation \cite{AL}. 
  
Having introduced $\hat{e}_{1}$ and $\hat{e}_{2}$ as two mutually orthogonal directions (both perpendicular 
to the direction of propagation of the radiation), and recalling the standard definitions of the Stokes parameters \cite{chandra} it can be easily 
shown that 
\begin{equation}
I= |\vec{E}\cdot\hat{e}_{1}|^2 +  |\vec{E}\cdot\hat{e}_{2}|^2,\qquad 
  V = 2 \,\mathrm{Im}[ (\vec{E}\cdot\hat{e}_{1})^{*} (\vec{E}\cdot\hat{e}_{2})],
 \label{one}
\end{equation}  
are both invariant for a rotation of $\hat{e}_{1}$ and $\hat{e}_{2}$ 
on the plane orthogonal to the direction of propagation of the radiation.
For the same two-dimensional rotation, $(Q\pm i U)$ transform as 
a function of spin weight $\mp 2$ on the two-sphere \cite{NP}; this observation leads, after some 
algebra, to the known form of the E-mode and B-mode polarization \cite{EB}.
If a large-scale magnetic field is present and if, concurrently, the spatial curvature 
does fluctuate over large scales, then the power spectra 
associated with the brightness perturbations of V will not be vanishing and shall be defined, in what follows, V-mode power spectra in analogy with the B-mode and E-mode power spectra characterizing the linear polarizations. 

To compute the induced V-mode polarization the evolution of the brightness perturbations 
must be written in the case when the photons scatter electrons in a magnetized 
background.  In the elastic $\mathrm{e}$-$\gamma$ scattering occurring in a cold plasma 
the recoil energy of the electron can be neglected \cite{BER}; photons impinging on electrons and ions in a weakly magnetized medium
can be described, as usual, in terms of a scattering matrix connecting the outgoing to the 
ingoing Stokes parameters (see, e.g. \cite{chandra}). The latter step will 
lead, after angular integration, to the evolution of the various brightness perturbations\footnote{In what follows 
$\epsilon' = a x_{\mathrm{e}} \tilde{n}_{0} \sigma_{\gamma\mathrm{e}}$ is the differential optical depth, $\tilde{n}_{0} = n_{0}/a^3$
is the physical concentration and $\sigma_{\gamma \mathrm{e}}= (8/3) \pi (e^2/m_{\mathrm{e}})^2$. The quantity $\mu$ is simply the projection 
of the Fourier wavevector on the direction of the photon momentum.}
\begin{eqnarray}
&& \Delta_{\mathrm{I}}' + i k\mu(\Delta_{\mathrm{I}} + \phi) + \epsilon' \Delta_{\mathrm{I}} = \psi' + \epsilon' {\mathcal C}_{\mathrm{I}}(\omega,k,\mu),
\label{DI}\\
&&  \Delta_{\mathrm{P}}' + i k \mu \Delta_{\mathrm{P}}  + \epsilon' \Delta_{\mathrm{P}} =  \epsilon' {\mathcal C}_{\mathrm{P}}(\omega,k,\mu),
\label{DP}\\
&& \Delta_{\mathrm{V}}' + ik \mu\Delta_{\mathrm{V}} + \epsilon' \Delta_{\mathrm{V}} =\epsilon' {\mathcal C}_{\mathrm{V}}(\omega,k,\mu),
\label{DV}
\end{eqnarray}
where the prime denotes a derivation with respect to the conformal time 
coordinate $\tau$ while 
$\phi = \delta_{\mathrm{s}}^{(1)} g_{00}/(2 a^2)$ and $\psi \delta_{ij} = \delta_{\mathrm{s}}^{(1)} g_{ij}/(2 a^2)$ are the scalar fluctuations of the metric whose relation to the curvature fluctuations can be expressed, in the longitudinal gauge, as 
\begin{equation}
{\mathcal R} = - \psi - \frac{{\mathcal H}^2}{{\mathcal H}^2 - {\mathcal H}'}\biggl( \phi 
+\frac{\psi'}{{\mathcal H}}\biggr), \qquad {\mathcal H} = \frac{a'}{a};
\label{R}
\end{equation}
note that the relation of ${\mathcal H}$ to the Hubble parameter is simply given by $a H= {\mathcal H}$; 
the source functions appearing in Eqs. (\ref{DI}), (\ref{DP}) and (\ref{DV}) are given by
\begin{eqnarray}
&& {\mathcal C}_{\mathrm{I}}(\omega,k,\mu) =
 \frac{1}{4}\biggl\{\Delta_{\mathrm{I}0} \biggl[ 2 \Lambda_{3}(\omega) (1 - \mu^2) + 2 \zeta^2(\omega) \biggl(\mu^2 
 + \Lambda_{1}^2(\omega) +  f_{\mathrm{e}}^2(\omega) \Lambda_{1}^2(\omega) (1 + \mu^2)\biggr)\biggr]
 \nonumber\\
 && + \biggl[ 2 \Lambda_{3}(\omega) (1 - \mu^2) - \zeta^2(\omega) \biggl(\mu^2 + \Lambda_{1}^2(\omega)\biggr) 
 - f_{\mathrm{e}}^2(\omega) \zeta^2(\omega) \Lambda_{2}^2(\omega) (1 + \mu^2)\biggr] S_{\mathrm{P}} + 4\mu v_{\mathrm{b}}
 \nonumber\\
 && - 6\, i \,f_{\mathrm{e}}^2(\omega) \zeta^2(\omega) \Lambda_{2}(\omega) \biggl[\mu^2 + \Lambda_{1}(\omega)\biggr]\Delta_{\mathrm{V}1}\biggr\},
 \label{CIF}\\
&&{\mathcal C}_{\mathrm{P}}(\omega,k,\mu) = \frac{1}{4}\biggl\{ \biggl[ 2 (1 - \mu^2)\biggl(\Lambda_{3}(\omega) - \zeta^2(\omega) f_{\mathrm{e}}^2(\omega) \Lambda_{2}^2(\omega)\biggr) - 2 \zeta^2(\omega) \biggl(\Lambda_{1}(\omega) -\mu^2 \biggr)\biggr]\Delta_{\mathrm{I}0}
 \nonumber\\
 &&+ \biggl[ 2 \Lambda_{3}(\omega) (1 - \mu^2) - \zeta^2(\omega) \biggl(\mu^2 - \Lambda_{1}^2(\omega)
 - f_{\mathrm{e}}^2(\omega) \Lambda_{2}^2(\omega) (1 - \mu^2)\biggr)\biggr] S_{\mathrm{P}}
 \nonumber\\
&&- 6 i f_{\mathrm{e}}^2(\omega) \zeta^2(\omega) \Lambda_{2}(\omega)\biggl(\mu^2 - \Lambda_{1}(\omega)\biggr)\Delta_{\mathrm{V}1}\biggr\},
 \label{CQF}\\
&&{\mathcal C}_{\mathrm{V}}(\omega,k,\mu) = \frac{\zeta^2(\omega) P_{1}(\mu)}{2}\biggl\{f_{\mathrm{e}}(\omega) \Lambda_{2}(\omega) \biggl( \Lambda_{1}(\omega) + 1\biggr) \biggl[ 2 \Delta_{\mathrm{I}0} - S_{\mathrm{P}} \biggr] 
  \nonumber\\
&& - \frac{3}{2} i \biggl[ \Lambda_{1}(\omega) + f_{\mathrm{e}}^2(\omega) \Lambda_{2}^2(\omega)\biggr] \Delta_{\mathrm{V}1}\biggr\},
\label{CVF}
 \end{eqnarray}
where the dependence upon the frequency of the observational channel arises through the functions
\begin{eqnarray}
&& \Lambda_{1}(\omega) = 1 + 
\biggl(\frac{\omega^2_{\mathrm{p\,\,i}}}{\omega^2_{\mathrm{p\,\,e}}}\biggr) \biggl( 
\frac{ \omega^2 - \omega^2_{\mathrm{B\,\,e}}}{\omega^2 - \omega^2_{\mathrm{B\,\,i}}}\biggr),\,\,\,\,
\Lambda_{2}(\omega) = 1 - \biggl(\frac{\omega^2_{\mathrm{p\,\,i}}}{\omega^2_{\mathrm{p\,\,e}}}\biggr)
\biggl(\frac{\omega_{\mathrm{B\,\,i}}}{\omega_{\mathrm{B\,\,e}}}\biggr) \biggl( 
\frac{ \omega^2 - \omega_{\mathrm{B\,\,e}}^2}{\omega^2 - \omega^2_{\mathrm{B\,\,i}}}\biggr),
\nonumber\\
&& \Lambda_{3}(\omega) = 1 + \biggl(\frac{\omega^2_{\mathrm{p\,\,i}}}{\omega^2_{\mathrm{p\,\,e}}}\biggr),\,\,\,\,
\zeta(\omega) = \frac{1}{f_{\mathrm{e}}^2(\omega) -1} = \frac{\omega^2}{\omega_{\mathrm{Be}}^2 - \omega^2}, \,\,\,\,\,
f_{\mathrm{e}}(\omega) =  \biggl(\frac{\omega_{\mathrm{B\,\,e}}}{\omega}\biggr).
\label{defmod2}
\end{eqnarray}
 The plasma and Larmor frequencies for electrons and ions are denoted, respectively, by 
($\omega_{\mathrm{p\,\,e}}$, $\omega_{\mathrm{B\,\,e}}$) and  ($\omega_{\mathrm{p\,\,i}}$, 
$\omega_{\mathrm{B\,\,i}}$).
 CMB experiments operate for angular frequencies which are 
typically larger than the Larmor\footnote{ 
Note that $\omega_{\mathrm{p\,e}}/\omega_{\mathrm{p\,i}} = \sqrt{m_{\mathrm{p}}/m_{\mathrm{e}}}$ where $m_{\mathrm{p}}$ is the proton mass; $\omega= 2 \pi \nu$ denotes throughout the angular frequency.} and plasma 
frequencies of the electrons at recombination; it is therefore 
legitimate to expand the  source functions of Eqs. (\ref{DI}), (\ref{DP}) and (\ref{DV}) 
in powers of $f_{\mathrm{e}}(\omega) = (\omega_{\mathrm{B\,e}}/\omega)$ 
as well as in powers of $(m_{\mathrm{e}}/m_{\mathrm{p}})$; the result 
of this double expansion can be written as
\begin{eqnarray}
&& {\mathcal C}_{\mathrm{I}}(\omega,k,\mu) = \biggl[\Delta_{\mathrm{I}0} + \mu v_{\mathrm{b}} - \frac{P_{2}}{2} S_{\mathrm{P}}\biggr]
\nonumber\\
&& + [ 2 + P_{2}(\mu)]f_{\mathrm{e}}^2(\omega)\biggl[ \Delta_{\mathrm{I}0} - i \Delta_{\mathrm{V}1} -\frac{S_{\mathrm{P}}}{2} \biggr]+ {\mathcal O}\biggl(\frac{m_{\mathrm{e}}}{m_{\mathrm{p}}}\biggr) + {\mathcal O}(f_{\mathrm{e}}^4),
\label{CexI1}\\
&& {\mathcal C}_{\mathrm{P}}(\omega,k,\mu) = \frac{1 -P_{2}(\mu)}{2} \biggl\{S_{\mathrm{P}} + 
f_{\mathrm{e}}^2(\omega)\biggl[ 2 i \Delta_{\mathrm{V}1}  - 2 \Delta_{\mathrm{I}0}  + S_{\mathrm{P}}\biggr]\biggr\} 
\nonumber\\
&&+ {\mathcal O}\biggl(\frac{m_{\mathrm{e}}}{m_{\mathrm{p}}}\biggr) + 
{\mathcal O}(f_{\mathrm{e}}^4),
\label{CexQ1}\\
&&{\mathcal C}_{\mathrm{V}}(\omega,k,\mu) = \frac{P_{1}(\mu)}{2} \biggl\{ 2 f_{\mathrm{e}}(\omega) 
\biggl[ 2 \Delta_{\mathrm{I}0} - S_{\mathrm{P}}\biggr] - \frac{3}{2} i 
\biggl[ 1 + f_{\mathrm{e}}^2(\omega)\biggr] \Delta_{\mathrm{V}1}\biggr\}
\nonumber\\
&&+ 
{\mathcal O}\biggl(\frac{m_{\mathrm{e}}}{m_{\mathrm{p}}}\biggr) + 
{\mathcal O}(f_{\mathrm{e}}^4),
\label{CexV1}
\end{eqnarray}
where $P_{\ell}(\mu)$ is the Legendre polynomial of $\ell$-th order and 
where $S_{\mathrm{P}}(k,\tau) = (\Delta_{\mathrm{P}0} + \Delta_{\mathrm{P}2} + 
\Delta_{\mathrm{I}2})$ is the standard source term for the E-mode 
polarization when $f_{\mathrm{e}}(\omega)=0$. Indeed, in the limit $f_{\mathrm{e}}(\omega) \to 0$, Eqs. (\ref{DI}), (\ref{DP}) and (\ref{DV}) reproduce the standard results for the evolution equations of the scalar brightness perturbations. The source functions obtained in Eqs. 
(\ref{CexI1})--(\ref{CexV1}) are derived by integrating the angular dependence of the ingoing Stokes parameters 
in full analogy with what happens in the case when the magnetic field is absent \cite{chandra} (see also \cite{MGR} 
for further details as well as \cite{MTR} for slightly different perspectives on magnetized photon-electron scattering). Note that  
\begin{equation}
f_{\mathrm{e}}(\omega) = \frac{\omega_{\mathrm{Be}}}{\omega}
 = 2.79 \times 10^{-12} \biggl(\frac{B_{\mathrm{u}}}{\mathrm{nG}}\biggr)
 \biggl(\frac{\mathrm{GHz}}{\nu}\biggr) (z_{\mathrm{rec}} +1),\qquad \omega_{\mathrm{Be}}=
 \frac{e  |\vec{B}\cdot\hat{n}|}{m_{\mathrm{e}}a},
 \label{FE}
 \end{equation}
 where $\omega_{\mathrm{B\, e}}$ is the Larmor frequency and $B_{\mathrm{u}}$ denotes the uniform 
 component of the comoving  magnetic field intensity which is treated within the guiding centre approximation 
 (see \cite{MGA} for the description of magnetized plasma prior to recombination).
The numerical solution of the system is greatly helped by exploiting systematically the integration along the line of sight \cite{EB} for all the brightness perturbations. From Eq. (\ref{DV}) 
\begin{equation}
\Delta_{\mathrm{V}}(k,\mu, \omega,\tau_{0}) = \int_{0}^{\tau_{0}} {\mathcal K}(\tau) \, P_{1}(\mu)\, \biggl\{ \, f_{\mathrm{e}}(\omega) \biggl[ 2\Delta_{\mathrm{I}0} - S_{\mathrm{P}} \biggr]  - \frac{3\, i}{4}   \Delta_{\mathrm{V}1}\biggr\} \,e^{-i \mu k (\tau_{0} - \tau)} d\tau,
\label{LS0}
\end{equation}
where ${\mathcal K}(\tau)$ is the visibility function and were $(\tau_{0}-\tau)$ is effectively the (comoving) angular diameter distance in a spatially flat geometry.  
To zeroth order in the tight-coupling expansion Eq. (\ref{LS0}) allows to evaluate the V-mode polarization, i.e. 
\begin{equation}
\Delta_{\mathrm{V}}(k,\mu,\tau_{0}) = \frac{8}{3}  \int_{0}^{\tau_{0}} {\mathcal K}(\tau) f_{\mathrm{e}}(\omega)
e^{- i \mu k (\tau_{0} - \tau)} \, P_{1}(\mu)\, \overline{\Delta}_{\mathrm{I}0}(k,\tau) \, d\tau.
\label{LS1}
\end{equation}
where $\overline{\Delta}_{\mathrm{I}0}(k,\tau)$ is the monopole of the intensity computed to lowest 
order in the tight coupling expansion, i.e. when the baryon velocity coincides with the dipole of the intensity 
of the radiation field. To lowest order in the 
tight coupling approximation the CMB is circularly polarized provided a large-scale magnetic field is 
present. The linear polarization is generated to first-order in the tight-coupling expansion but is larger 
than the V-mode polarization because of the smallness of $f_{\mathrm{e}}(\omega)$. More details
on this semi-analytic discussion can be found in \cite{MGR}.
The evolution of the monopole of the intensity $\Delta_{\mathrm{I}0}$ can either 
be studied in the tight-coupling limit or it can be solved numerically. In both cases
the initial conditions will be chosen to be adiabatic \footnote{See, e.g., \cite{BER}. The present 
considerations can be straightforwardly generalized to the case of non-adiabatic initial conditions \cite{nad}. For a discussion on the peculiar features of the adiabatic and non-adiabatic initial 
conditions of the Einstein-Boltzmann hiererchy see, for instance, \cite{primer}.}
\begin{equation}
\Delta_{\mathrm{I}0}(k,\tau_{\mathrm{rec}})= \frac{ 2 ( R_{\nu} + 15)}{5 ( 4 R_{\nu} + 15)} {\mathcal R}_{*}(k),\qquad  \psi_{*}(k) = \biggl( 1 + \frac{2}{5} R_{\nu} \biggr) \phi_{*}(k),
\end{equation}
where $R_{\nu}$ is the fractional contribution of the massless 
neutrinos to the radiation background and ${\mathcal R}_{*}(k)$ denotes 
the curvature perturbations prior to equality and for typical scales 
larger than the Hubble radius at the corresponding epoch. Since 
prior to equality ${\mathcal H} = 1/\tau$ we shall also have, from Eq. 
(\ref{R}), that ${\mathcal R}_{*}(k) = - \psi_{*}(k) - \phi_{*}(k)/2$.
For large angular scales (i.e. $\ell < 50$) the visibility function can be considered 
to be sharply peaked at recombination and it is in practice a Dirac delta 
function. The V-mode autocorrelation (i.e., for short, VV power spectrum) and the 
cross-corrrelation between polarization and temperature  (i.e., for short, VT power spectrum)
can then be computed analytically in this regime and the result \footnote{For simplicity the angular power spectrum shall be denoted as $G_{\ell}^{(\mathrm{XY})} = \ell(\ell +1) C_{\ell}^{(\mathrm{XY})}/(2\pi)$.
The VV and VT power spectra are the analog of the EE and TE power spectra arising in the case of the 
linear polarization. The temperature (related to the I Stokes parameter)  and the circular polarization (related to the V- Stokes parameter) are both invariant under a rotation orthogonal to the 
direction of propagation of the radiation, as stressed after Eq. (\ref{one}). It is therefore natural, in a first approach to the problem, to consider the TT,  VT 
and the VV power spectra. Furthermore, the VT correlations are larger than the the VE and VB correlations.} can be written as:
\begin{eqnarray}
&& G_{\ell}^{(\mathrm{VV})}(\omega) =  \frac{256 \pi}{225} \biggl(\frac{R_{\nu} + 15}{4 R_{\nu} + 15}\biggr)^2 f_{\mathrm{e}}^2(\omega) {\mathcal A}_{{\mathcal R}} \biggl(\frac{k_{0}}{k_{\mathrm{p}}}\biggr)^{n_{\mathrm{s}}-1} {\mathcal I}_{\ell}^{(\mathrm{VV})}(n_{\mathrm{s}}),
\label{VV}\\
&& G_{\ell}^{(\mathrm{VT})}(\omega) = 
\frac{16\pi}{75}\frac{(R_{\nu} + 15) ( 2 R_{\nu} - 15)}{(4 R_{\nu} + 15)^2}  f_{\mathrm{e}}(\omega) {\mathcal A}_{{\mathcal R}} 
\biggl(\frac{k_{0}}{k_{\mathrm{p}}}\biggr)^{n_{\mathrm{s}} -1} \,
{\mathcal I}_{\ell}^{(\mathrm{VT})}(n_{\mathrm{s}}),
\label{VT}
\end{eqnarray}
where $k_{0} = (\tau_{0}- \tau_{\mathrm{rec}})^{-1}$ and 
$k_{\mathrm{p}} =0.002\,\,\mathrm{Mpc}^{-1}$ is the pivot scale while 
${\mathcal A}_{{\mathcal R}}$ is the amplitude of the power spectrum of curvature perturbations at $k_{\mathrm{p}}$ (in the concordance 
paradigm and in the light of the WMAP data alone ${\mathcal A}_{{\mathcal R}}= (2.41 \pm 0.11) \times 10^{-9}$); the functions ${\mathcal I}_{\ell}^{(\mathrm{VV})}(n_{\mathrm{s}})$ and ${\mathcal I}_{\ell}^{(\mathrm{VT})}(n_{\mathrm{s}})$  appearing in Eqs. (\ref{VV}) and (\ref{VT}) are nothing but:
\begin{eqnarray}
&&{\mathcal I}_{\ell}^{(\mathrm{VV})}(n_{\mathrm{s}}) = 
\frac{\ell(\ell +1)[ 4 \ell (\ell +1) - (n_{\mathrm{s}} -1) (n_{\mathrm{s}}-2) (n_{\mathrm{s}}-4)] \Gamma(3 -n_{\mathrm{s}}) \Gamma\biggl(\ell - \frac{3}{2} + \frac{n_{\mathrm{s}}}{2}\biggr)}{2^{6 - n_{\mathrm{s}}} \Gamma\biggl(2 - \frac{n_{\mathrm{s}}}{2}\biggr)
\Gamma\biggl(3 - \frac{n_{\mathrm{s}}}{2}\biggr) \Gamma\biggl(\frac{7}{2} +\ell - 
\frac{n_{\mathrm{s}}}{2} \biggr)}
\nonumber\\
&& {\mathcal I}_{\ell}^{(\mathrm{VT})}(n_{\mathrm{s}}) =\frac{\ell (\ell +1)(2-n_{\mathrm{s}}) \Gamma\biggl(2-\frac{n_{\mathrm{s}}}{2}\biggr) 
\Gamma\biggl(\ell+\frac{n_{\mathrm{s}}}{2}-1\biggr)}{4 \sqrt{\pi } \Gamma\biggl(\frac{5}{2}-\frac{n_{\mathrm{s}}}{2}\biggr)
 \Gamma\biggl(\ell-\frac{n_{\mathrm{s}}}{2}+3\biggr)},
\label{VTint}
\end{eqnarray}
and arise as analytically solvable integrals of products of spherical Bessel functions and of their 
derivatives. Before discussing the relevant numerical results over small angular scales it is 
appropriate to mention here that circular polarization is often invoked as the result 
of the Faraday conversion\footnote{Faraday conversion (typical of relativistic jets) should not be confused with Faraday rotation. In the presence of relativistic electrons linearly polarized radiation can be Faraday converted into circularly polarized radiation \cite{SY}. Faraday 
rotation is instead a rotation of the polarization plane 
of the (linearly polarized) radiation: in practice it can convert E-modes into B-modes but it does not lead to circularly polarized photons (see, e.g.  third and last references of \cite{MGA}). Faraday conversion and Faraday rotation have also different dependences upon the magnetic field intensity and upon the frequency.} of linearly polarized radiation \cite{SY}. For the latter 
mechanism to operate, relativistic electrons must be present in the system and this can happen 
only as a secondary effect when CMB photons pass through magnetized clusters (see \cite{SY}, last reference); this is however not the idea pursued here since the pre-decoupling plasma is cold and electrons are deeply non-relativistic. The V-mode polarization, as we showed, is induced by the magnetized plasma itself thanks to the 
presence of (adiabatic) curvature perturbations in the system. Absent one of these two components 
the VT and VV power spectra would vanish. If the initial conditions would not be adiabatic 
the V-mode polarization would still be present but with different physical features which will depend upon the 
specific non-adiabatic solution \cite{nad,primer}.
For smaller angular scales (i.e. $\ell >100$) it is mandatory  to 
integrate numerically the system across decoupling.
\begin{figure}[!ht]
\centering
\includegraphics[height=6cm]{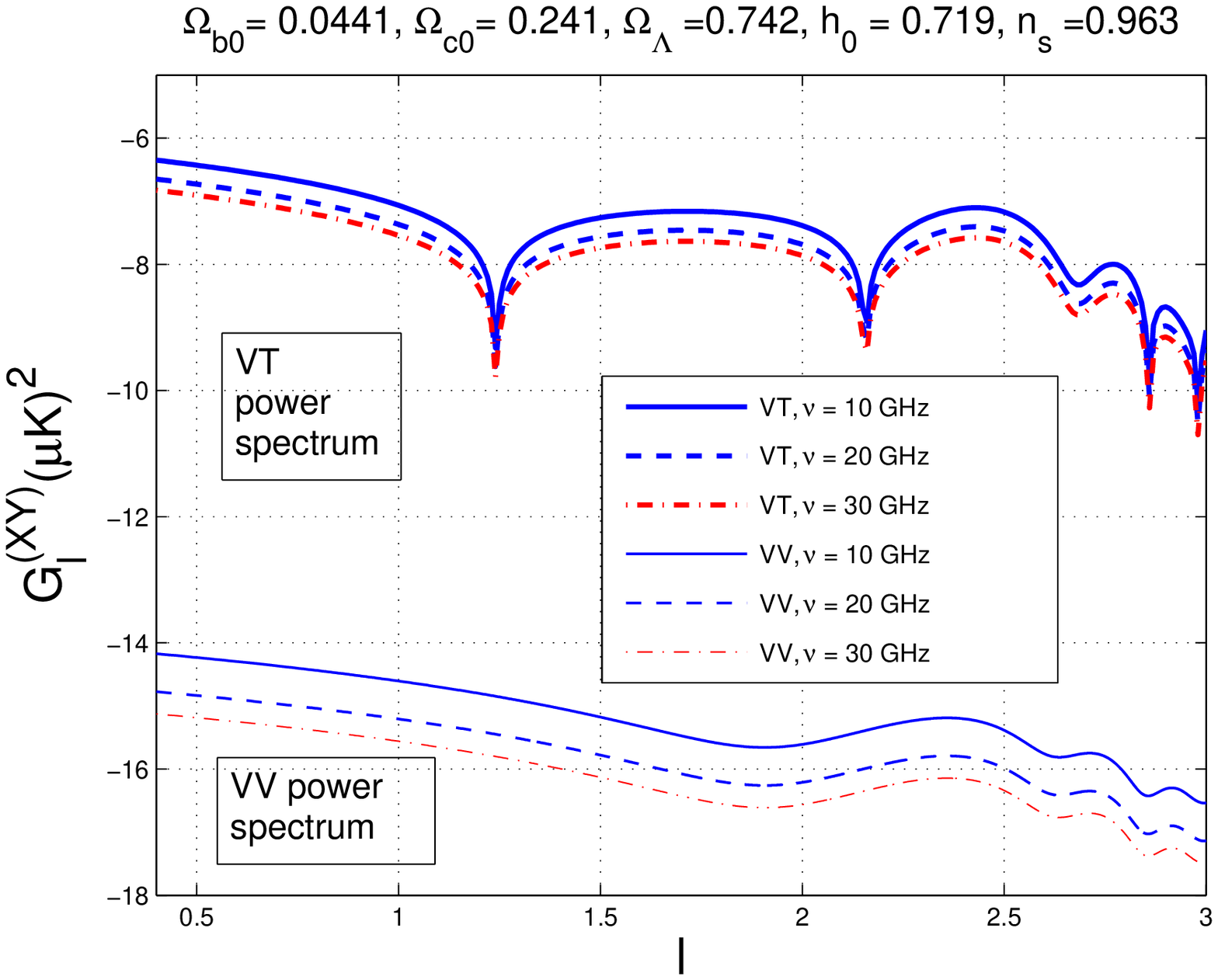}
\includegraphics[height=6cm]{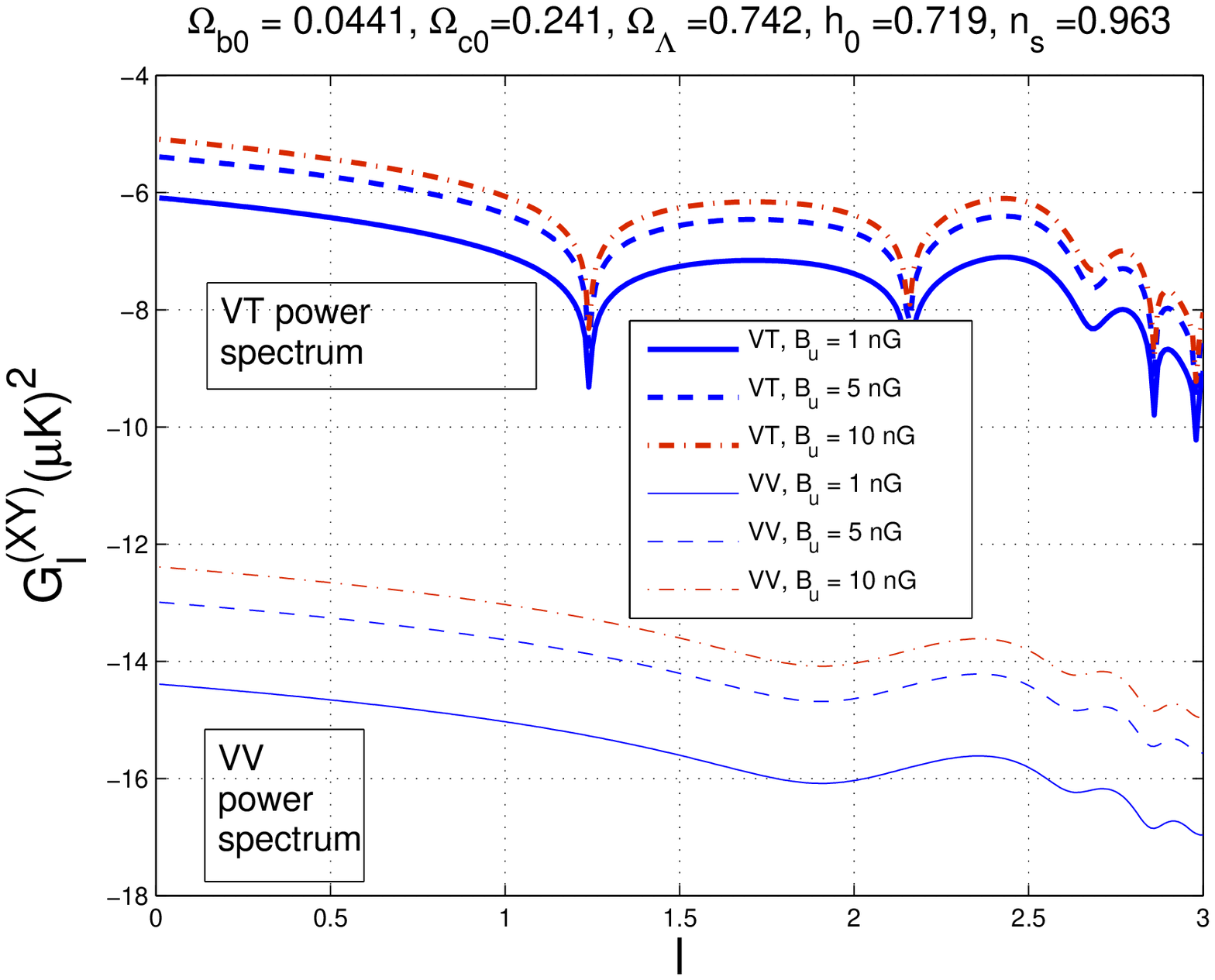}
\caption[a]{In the plot at the left the VT and the VV angular power spectra are reported 
for a fixed value of $B_{\mathrm{u}}$ (i.e. $1$ nG) but for different values of the 
comoving frequency. In the plot at the right the comoving frequency is fixed 
to $10$ GHz but the magnetic field strength increases. The thin lines denote 
the VV correlations while the thick lines denote the VT correlations. In the plots, on both axes, the 
common logarithm of the corresponding quantity is reported.}
\label{FIG1}      
\end{figure}
Some of the results are summarized in Fig. \ref{FIG1}. 
The thin lines in both plots  denote the V-mode autocorrelations while the thick lines denote 
the cross-correlation of the circular polarization anisotropies 
with the temperature inhomogeneities. 
The signal is larger for low multipoles and its 
shape reminds a bit of the temperature autocorrelations 
induced by the tensor modes of the geometry which reach their largest 
value for small $\ell$ and decay exponentially for $\ell > 90$.  Defining as $r_{\mathrm{T}}$ the tensor to scalar 
ratio at the pivot scale $k_{\mathrm{p}}$ \cite{WMAP5,primer}, for $r_{\mathrm{T}}=1$ 
the TT correlations induced by the tensor modes would be ${\mathcal O}(10^{3})\,(\mu\mathrm{K})^2$ while the VT correlations are 
${\mathcal O}(10^{-5})\,(\mu\mathrm{K})^2$ for the choice of parameters 
of Fig. \ref{FIG1} (see, for instance, plot at the right, dot-dashed curve). For the same choice of parameters the VT power spectra 
are of the order of the B-mode autocorrelation induced by the weak lensing 
of the primary anisotropies (i.e. ${\mathcal O}(10^{-5}) (\mu\mathrm{K})^2$, see also left plot of
 Fig. \ref{FIG2} and the discussion below). By shifting the observational frequency the VT correlation can be even larger \cite{GS,GS2}.
The B-mode autocorrelation induced by the tensor modes of the geometry 
is typically larger, both than the V-mode polarization and than the BB spectra from lensing. For $r_{\mathrm{T}} \sim 0.1$, the BB angular power spectrum of the tensor modes of the geometry is 
${\mathcal O}(10^{-2}) (\mu\mathrm{K})^2$ for $\ell \sim 90$ corresponding 
to angular separations of roughly $2$ deg. 
In Fig. \ref{FIG2} (plot at the left) the EE power spectrum stemming from the 
best fit to the WMAP 5-yr data alone is illustrated with a dashed line and compared, in the 
same plot, with  the BB angular power spectrum arising from the lensing 
of the primary anisotropies (thin dot-dashed line) as well as with the V-mode autocorrelation (full thin curve at the bottom). The B-mode autocorrelation stemming from 
the tensor modes in the case $r_{\mathrm{T}} =0.1$ is illustrated with 
the thick line.
\begin{figure}[!ht]
\centering
\includegraphics[height=6cm]{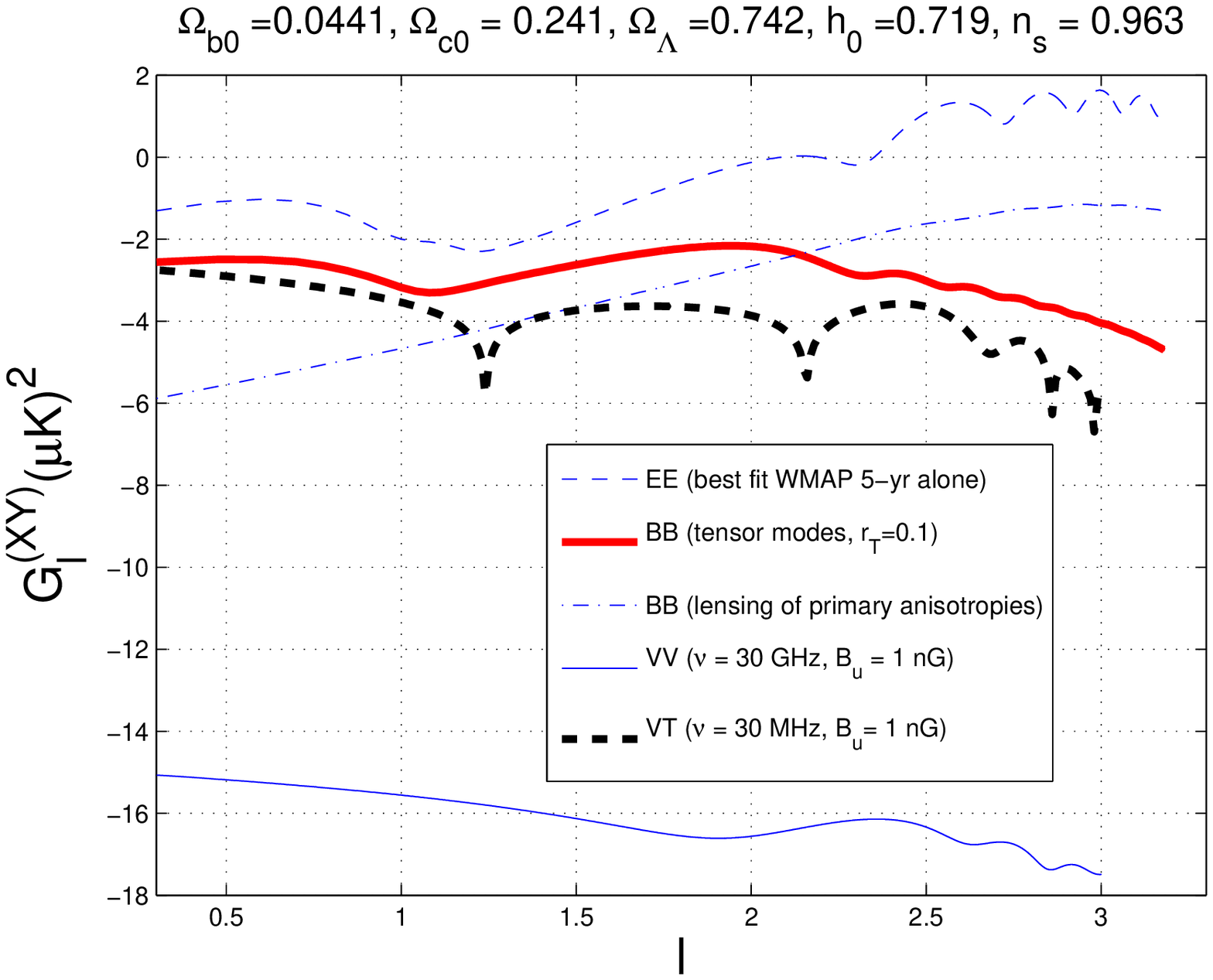}
\includegraphics[height=6cm]{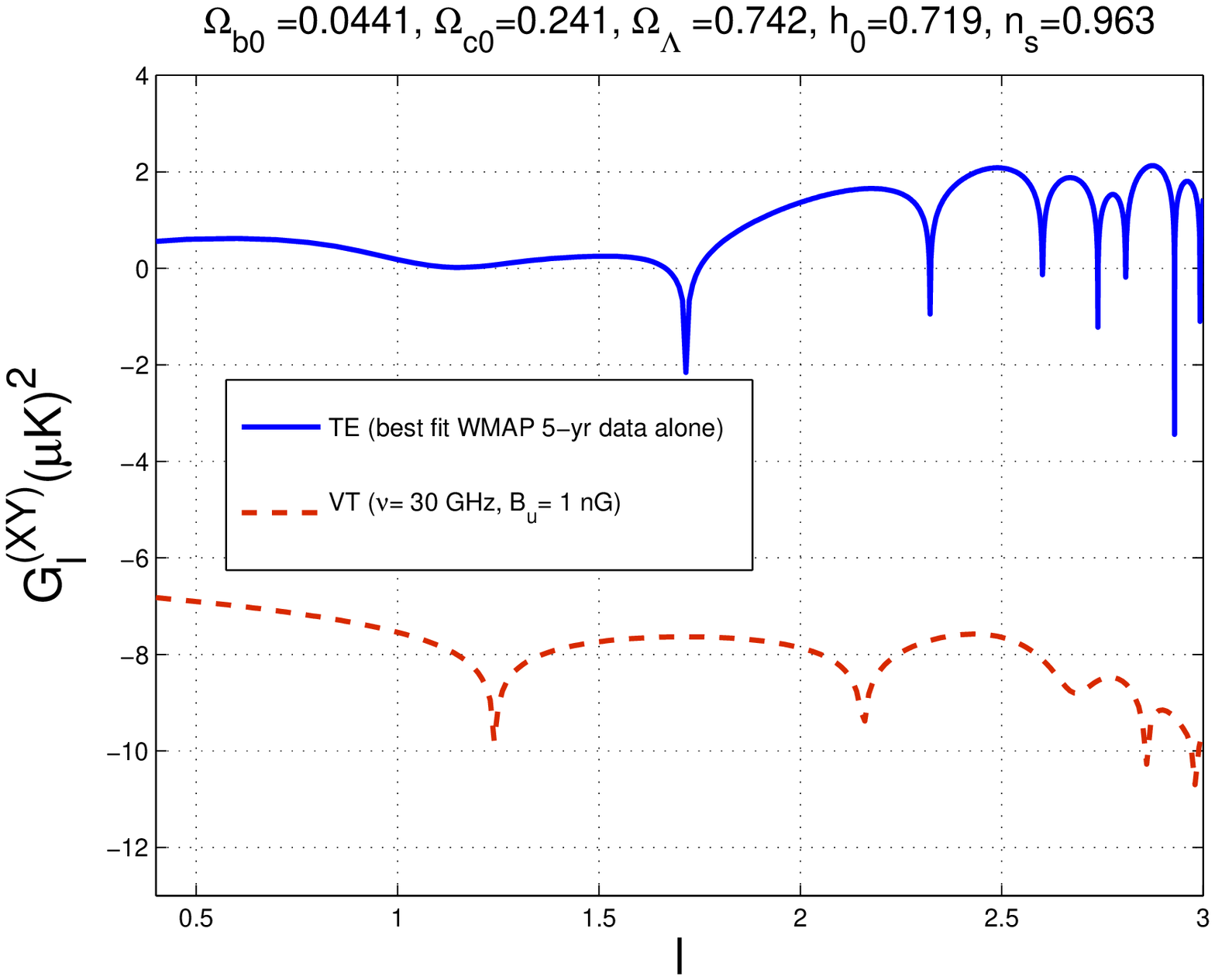}
\caption[a]{The V-mode power spectra are compared with the linear polarizations. 
As in Fig. \ref{FIG1} on both axes the common logarithm of the indicated quantity is reported.}
\label{FIG2}      
\end{figure}
Always in Fig. \ref{FIG2} (plot at the right) the TE and the VT correlations are compared.
Both in Figs. \ref{FIG1} and \ref{FIG2}
the cosmological parameters are fixed (as indicated in the title of each figure) to 
the values of the best fit stemming from the WMAP 5-yr data alone in the light of the concordance model.
In Fig. \ref{FIG2} the frequency of the channel has been taken of the order of $30$ GHz. 
Even if the latter frequency is already rather low, it would be 
desirable to reduce it even more and to conceive spectropolarimetric 
measurements in the range of the GHz.  The challenge of detecting  
the CMB radiation at low frequencies 
is neatly described in Ref. \cite{GS} where a set of absolute radiometers 
is employed in different channels at  $0.6$, $0.82$ and $2.5$ GHz (see also \cite{GS2} for earlier results 
along the same theme). 
As specifically discussed also in analytic terms (see Eqs. (\ref{VV})--(\ref{VT})) the VV and VT 
power spectra are sensitive to the underlying cosmological parameters, to the initial conditions 
of the Einstein-Boltzmann hierarchy as well as to the magnetic field parameters. For illustration 
the concordance model supplemented by adiabatic initial conditions has been considered. 
The maximal intensity of the comoving magnetic field has been taken to be of the order of the nG. This 
is the range of current bounds stemming from the simultaneous analysis of the measured TE and TT power 
spectra (see \cite{MGA}, first and second reference). Larger magnetic fields would distort the acoustic oscillations in the TT power spectra. 
Low frequency instruments 
could make the difference for scrutinizing a potential V-mode polarization. 
In this respect the results and the techniques of \cite{GS} (as well as the earlier results of \cite{GS2}) could be probably revisited in the light of the considerations 
developed here. It has been shown that the VT correlation for a comoving magnetic field 
from $5$ to $10$ nG can be as large as $10^{-5}\, (\mu \mathrm{K})^2$ at $10$ GHz for $\ell < 20$ (i.e. large angular separations). This means that for frequencies ${\mathcal O}(\mathrm{MHz})$, the resulting signal 
could be even $6$ or $7$ orders of magnitude larger than a putative B-mode signal from gravitational lensing (see, e.g. Fig. \ref{FIG2}, thick dashed curved in the left plot). 
It has been demonstrated that the study of circular dichroism is not more forlorn than other signals which are 
often invoked as conceptually important to consider but observationally difficult to assess. The systematic effects plaguing the measurements of the V-mode power spectra differ from the case of linear polarizations. Wether or not they are less severe depends also upon the features of the instrument and on the specific frequency band.  The author is grateful to G. Sironi, M. Gervasi and A. Tartari  for stimulating discussions.

\end{document}